# ANÁLISE TÉRMICA-FLUIDODINÂMICA DE UM TRANSFORMADOR DE ENERGIA "A SECO" ATRAVÉS DA DINÂMICA DOS FLUIDOS COMPUTACIONAL

| **Anderson Santos Nunes(*)** | **Allan Schwanz** | **Eduardo Postali** | **Marcelo Kruger** |
| WEG T&D | WEG T&D | WEG T&D | ESSS |
| **Unidade Blumenau** | **Unidade Blumenau** | **Unidade Blumenau** | |

RESUMO

O sistema de refrigeração é um fator fundamental nos projetos de transformadores de energia. A freqüente necessidade por projetos otimizados aumenta ainda mais a importância do desenvolvimento através de ferramentas cada vez mais desenvolvidas.

Este trabalho tem como objetivo avaliar um projeto de transformador, no qual a refrigeração é realizada através de exaustores e trocadores de calor instalados na lateral menor do transformador. O modelo é elaborado aplicando-se uma simetria, reduzindo-se o modelo a ser analisado.

Para avaliar a melhor configuração em termos de posicionamento, vazão e quantidade de exaustores e entender as principais características térmicas e fluidodinâmicas do escoamento no interior do transformador, utiliza-se a ferramenta de Dinâmica dos Fluidos Computacional. A metodologia empregada consiste em resolver o problema numericamente, através do pacote computacional ANSYS CFX®, as equações do escoamento e energia no interior do tanque do transformador.

PALAVRAS-CHAVE

Transformador, Dinâmicas dos Fluidos Computacional, refrigeração.

1. INTRODUÇÃO

A eficiência da refrigeração em um transformador é um fator fundamental que determina a segurança operacional e o tempo de vida do transformador. Atualmente alguns sistemas são bastante utilizados, entre eles o de refrigeração natural onde o calor é absorvido pelo ar e dissipado ao ambiente. Entretanto, nesses casos, faz-se necessário um grande volume de ar para que seja possível a dissipação de grandes valores de perdas. Porém, é cada vez mais freqüente a necessidade por projetos menores, que possam ser instalados em espaços delimitados, como por exemplo, em navios. Dessa forma, visando avaliar um projeto de transformador com razoável grau de otimização, foi desenvolvido um modelo de transformador, inerte ao ar ambiente, com trocadores de calor ar-água e com exaustores.

Para determinar a melhor configuração em termos de posicionamento, vazão e quantidade de exaustores e entender as principais características térmicas e fluidodinâmicas do escoamento no interior deste, foi desenvolvido um modelo numérico através do pacote computacional ANSYS CFX.

Primeiramente foi desenvolvido um modelo computacional contendo dois exaustores posicionados de maneira simétrica nos lados do tanque. Em seguida, para efeito de comparação, foram adicionados mais um exaustor de cada lado.

(*) Rua Dr. Pedro Zimmermann, n° 6751, CEP 89.068-001, Blumenau, SC, Brasil
Tel: (+55 47) 3337-1000 – Fax: (+55 47) 3337-1090 – Email: santos@weg.net

Os modelos permitiram avaliar a influência dos dispositivos internos do transformador no perfil térmico e fluidodinâmico do escoamento.

Os resultados mostraram um aumento significativo nos valores de velocidade no interior do transformador, em especial no interior dos canais entre os enrolamentos com a utilização de mais exaustores. Entretanto, a diferença no perfil térmico do ar na saída para os exaustores ficou em torno somente de 5°C entre os modelos.

## 2. TRANSFORMADOR REAL E O MODELO NUMÉRICO

A Figura 1 mostra o modelo numérico contendo dois exaustores elaborado para a simulação, bem como a descrição dos componentes pertencentes ao modelo. Observa-se a complexidade da geometria.

Os núcleo magnético e os enrolamentos são considerados como fonte de calor, de forma homogênea no volume. Os valores de condutividade térmica nas duas dimensões (axial e radial) foram modelados analiticamente.

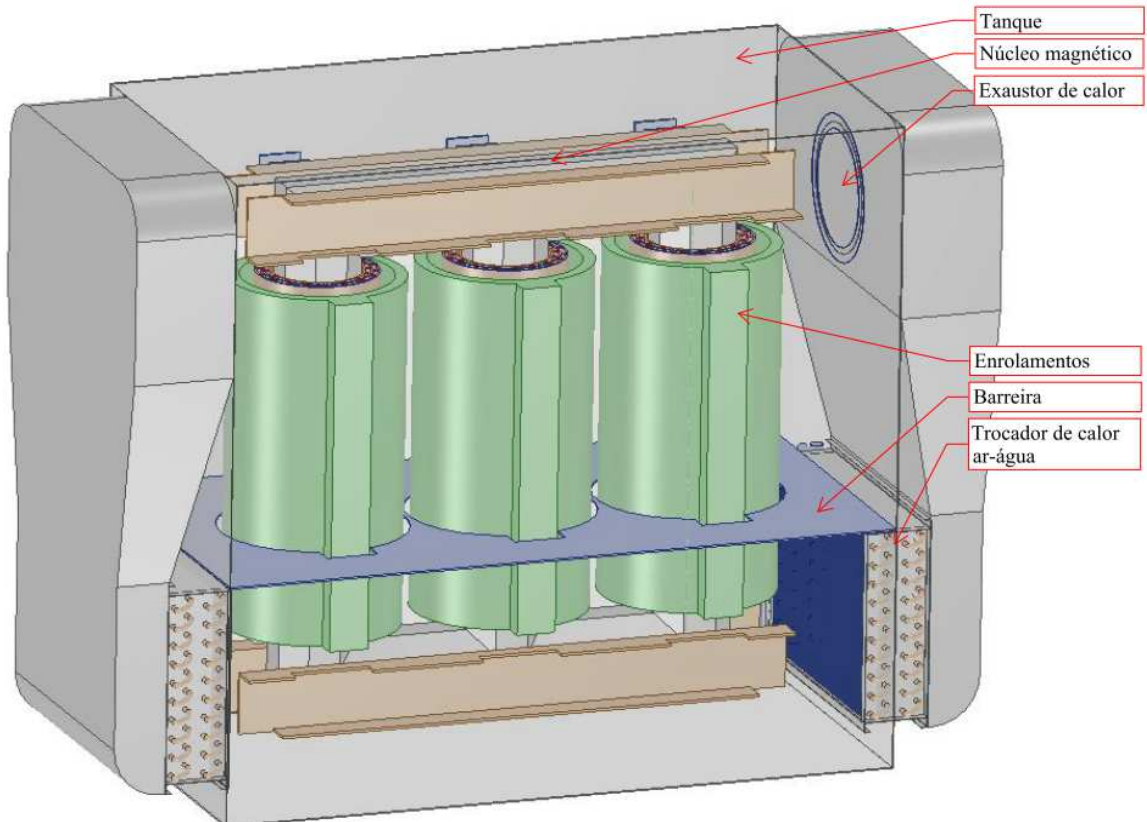

FIGURA 1 - Modelo numérico.

## 3. MODELAGEM NUMÉRICA

Os modelos numéricos para a solução de problemas envolvendo mecânica dos fluidos devem descrever os aspectos mais importantes da física do problema real. O domínio de cálculo em questão inclui o interior de um transformador a seco composto por 3 conjuntos de enrolamentos e núcleo magnético. Para modelar o escoamento no interior deste domínio deve-se lançar mão das equações de Navier-Stokes. Adicionalmente, condições de contorno para cada uma das variáveis independentes do problema devem ser fornecidas em cada fronteira do domínio.

*3.1 Equações Governantes do Escoamento*

Escoamentos de fluidos são regidos pelas equações de Navier-Stokes, mas a extrema complexidade destas equações impossibilita, até mesmo para casos simples, a obtenção de uma solução, mesmo com os recursos computacionais atualmente disponíveis, se uma simulação considerando todas as escalas espaciais e temporais (transientes) for desejada. Para modelar estes escoamentos são utilizados modelos aproximados, tais como as equações de Navier-Stokes com valores médios no tempo, conhecidas como Reynolds Averaged Navier-Stokes Equations (RANS) e que são empregadas neste trabalho.

Tomando-se a média das equações de Navier-Stokes surgem novos termos, conhecido por tensões de Reynolds, que carregam as características turbulentas e transitórias do escoamento. Uma das metodologias mais usuais consiste em modelar este tensor a partir das variáveis k (energia cinética turbulenta) e ε (taxa de dissipação turbulenta).

A equação da continuidade, independente do tempo, é dada por:

$$\nabla \cdot \mathbf{V} = S_{massa} \qquad (1)$$

onde $\mathbf{V}$ é o vetor velocidade médio no tempo de cada fluido $\{\mathbf{V_x}\,\mathbf{V_y}\,\mathbf{V_z}\}$. O termo $S_{massa}$ é o termo fonte de geração ou retirada de massa. Neste problema o termo fonte é nulo.

A equação de conservação da quantidade de movimento é dada por:

$$\nabla \cdot ((\rho \mathbf{V} \otimes \mathbf{V})) = \nabla \left( \mu_{eff} \left( \nabla \mathbf{V} + (\nabla)^T \right) \right) - \nabla \mathbf{p} + S_{QM} \qquad (2)$$

onde $\rho$ é a densidade, $\mathbf{p}$ é a pressão, e $\mathbf{S}_{QM} = \{\mathbf{S}_{QM_x}, \mathbf{S}_{QM_y}, \mathbf{S}_{QM_z}\}$ é o termo fonte de geração de quantidade de movimento.

Para a solução dos parâmetros turbulentos k e $\varepsilon$ é utilizado o modelo de turbulência $k - \varepsilon$ padrão. As equações de transporte são:

$$\nabla((\rho \mathbf{K} \mathbf{V})) = \nabla \left( \left( \mu_0 + \frac{\mu_t}{\Pr_K} \right) \nabla \mathbf{K} \right) + \Phi - \rho \varepsilon \qquad (3)$$

$$\nabla((\rho \varepsilon \mathbf{V})) = \nabla \left( \left( \mu_0 + \frac{\mu_t}{\Pr_\varepsilon} \right) \nabla \varepsilon \right) + C_1 \frac{\varepsilon}{\mathbf{K}} \Phi + C_2 \rho \frac{\varepsilon^2}{\mathbf{K}} \qquad (4)$$

onde $\Phi_{aço}$ é definido como

$$\Phi = \mu_{eff} \nabla \mathbf{V} \left( \nabla \mathbf{V} + (\nabla \mathbf{V})^T \right) - \frac{2}{3} \nabla \cdot \mathbf{V} \left( \mu_{eff} \nabla \cdot \mathbf{V} + \rho \mathbf{K} \right) \qquad (5)$$

e a viscosidade efetiva $\mu_{eff}$ é

$$\mu_{eff} = \mu_0 + \mu_t \qquad (6)$$

onde $\mu_0$ é a viscosidade molecular e $\mu_t$ é calculado a partir dos parâmetros turbulentos:

$$\mu_t = C_\mu \rho \frac{K^2}{\varepsilon} \qquad (7)$$

No caso de problemas que envolvam o cálculo de temperatura e transferência de calor, a equação de energia que rege o fenômeno é dada por:

$$\nabla \cdot (\rho \mathbf{V} h_{tot}) = \nabla \cdot (\lambda \nabla T) + S \qquad (8)$$

onde $\lambda$ é a condutividade térmica, $h_{tot} = h + \frac{1}{2} \mathbf{U}^2$ é a entalpia total e $h = h(p,T)$ é a entalpia estática, calculada a partir das relações termodinâmicas. Para o caso de um fluido com densidade e calor específico constante, a entalpia é simplesmente o produto do calor específico e da temperatura:

$$h = C_P \cdot T \qquad (9)$$

## 3.2 Empuxo Térmico

Para pequenas variações de densidade em função da temperatura, a aproximação de Boussinesq por ser empregada. Essas pequenas variações de densidade é que dão origem aos movimentos de massa entre as regiões frias e quentes do aço líquido.

A força de empuxo térmico é introduzida na equação da quantidade de movimento como uma força de campo, $\mathbf{B}$:

$$\mathbf{B} = (\rho - \rho_{ref})\mathbf{g} \qquad (10)$$

onde $\mathbf{g}$ é o vetor gravidade $\{0, -9.81, 0\}$ e $\rho_{ref}$ é a densidade de referência. É utilizada a aproximação de Boussinesq, onde os efeitos da variação da compressibilidade do gás na densidade podem ser ignorados.

O termo $(\rho - \rho_{ref})$ é descrito por:

$$(\rho - \rho_{ref}) = -\rho_{ref}\beta(T - T_{ref}) \qquad (11)$$

onde $\beta$ é a Expansividade Térmica, em $[^0C^{-1}]$ e $T_{ref}$ é a temperatura de referência.

## 3.3 Malha Computacional

A etapa de geração de malha, ou discretização do domínio de cálculo é uma das etapas mais importantes do pré-processamento de uma análise de CFD e é com certeza uma das mais decisivas na qualidade das soluções numéricas obtidas.

Para a discretização do domínio em análise foram utilizadas malhas híbridas, compostas por elementos hexaédricos nas regiões das bobinas e enrolamentos e tetraédricos/primas nas regiões do escoamento de ar no interior do transformador. A utilização dos elementos prismáticos próximo às paredes se faz necessário para melhor caracterização da camada limite fluidodinâmica e térmica nas superfícies, ver Figura 2.

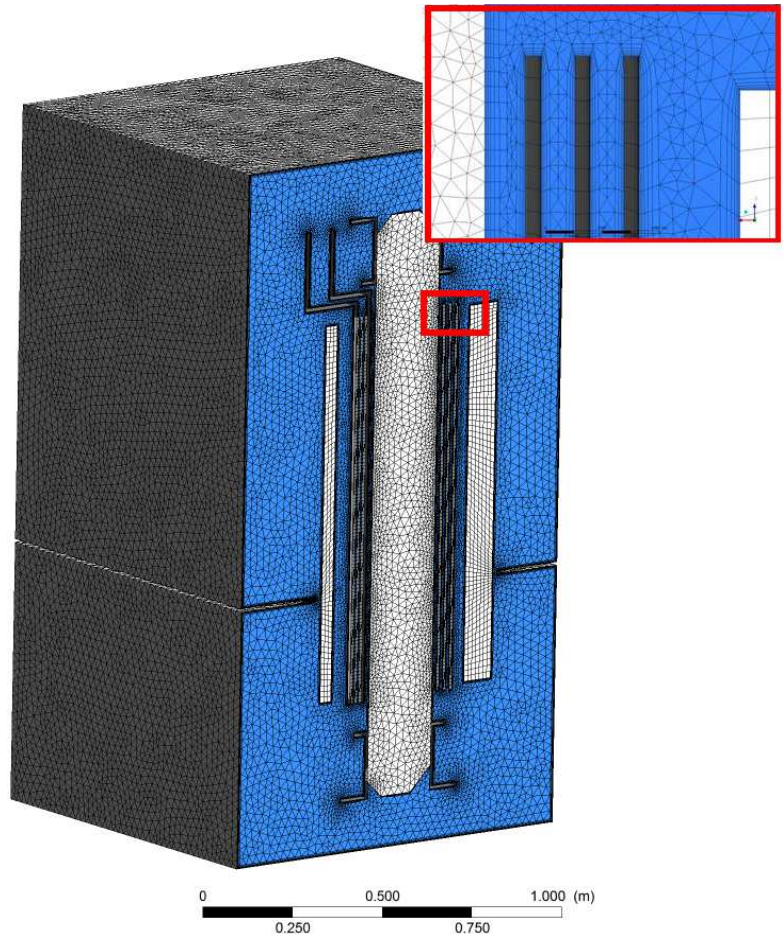

FIGURA 2 - Malha computacional.

*3.4 Condições de contorno e Simplificações*

As condições de contorno e principais considerações utilizadas no modelo numérico no presente trabalho são:

- Não foram considerados os sistemas laterais do exaustor e trocador de calor. Foi imposto um fluxo de ar de saída na face dos exaustores e entrada na face dos trocadores como condições de contorno;
- Foi utilizado a condição de simetria no centro (fase B) do transformador;
- O escoamento no interior do transformador é considerado tridimensional, turbulento, incompressível, com transferência de calor e monofásico (ar);
- Para a condição de contorno nas bobinas, enrolamentos e núcleo magnético foi utilizado fluxo de calor prescrito;
- Foi utilizado condição inicial de fluido estagnado com temperatura de 40ºC ;
- A função de interpolação utilizada para os termos advectivos foi o High Resolution;

## 4. RESULTADOS

A Figura 4 apresenta os perfis de temperatura no plano de simetria para os modelos de 2 (a) e 4 (b) exaustores. Primeiramente observa-se a uniformidade do perfil, temperatura uniforme em torno de 60 ºC, na região inferior dos dois transformadores. Esse comportamento deve-se a utilização da barreira colocada no centro do transformador que obriga com que o escoamento ocorra pelo interior das bobinas.
Na região superior é importante verificar que o modelo com 4 exaustores apresenta temperaturas menores entre os canais do enrolamento.

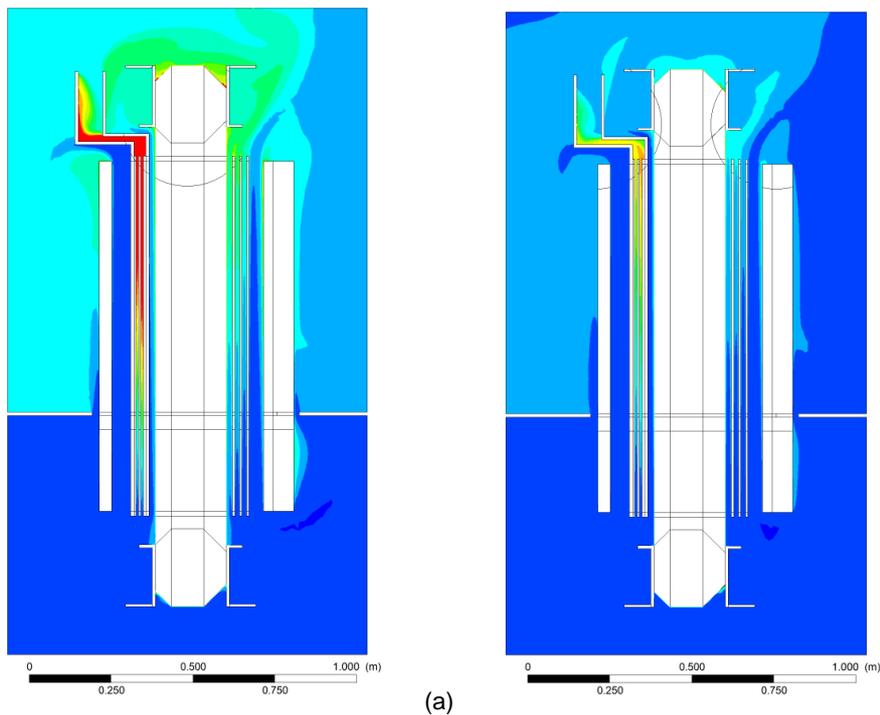

FIGURA 4 - Perfil de temperatura no plano de simetria. (a) modelo de 2 exaustores (b) modelo de 4 exaustores.

Na Figura 5, que representa o perfil de velocidade no plano de simetria, é possível comprovar o comportamento observado no plano térmico. A utilização da placa central promove as altas velocidades no interior das bobinas, melhorando a troca térmica nessas regiões.
Outra característica importante observada nos dois modelos é a influência das vigas superiores no perfil de velocidade. Observa-se regiões de recirculação e jatos direcionados à parede do tanque.

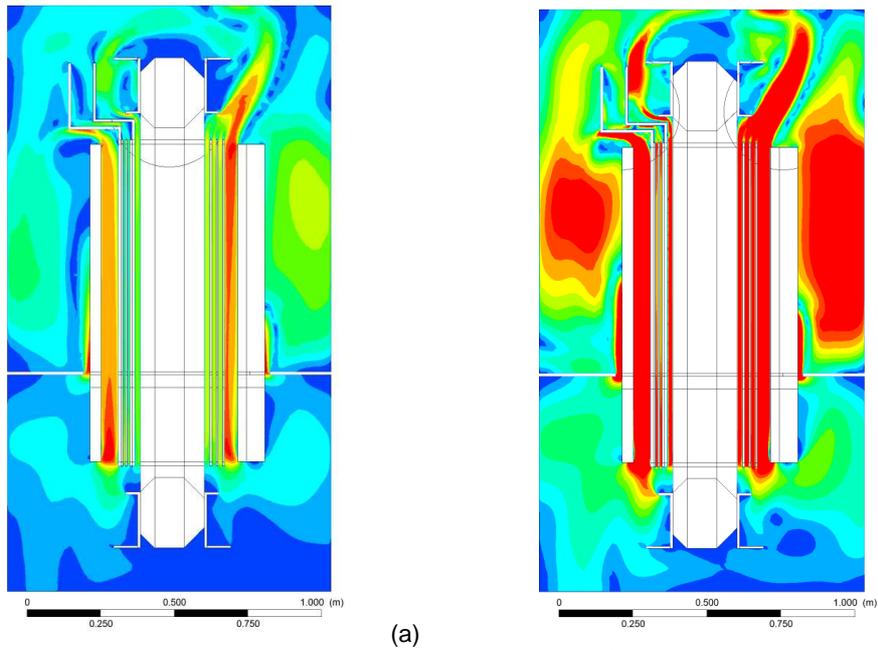

FIGURA 5 - Perfil de velocidade no plano de simetria. (a) modelo de 2 exaustores (b) modelo de 4 exaustores.

Por fim, a Figura 6 apresenta o mesmo perfil de temperatura porém considerando os componentes sólidos, enrolamentos e núcleo magnético. O aumento no número de exaustores promove um melhor resfriamento dos componentes internos do transformador.
Verifica-se uma diferença de aproximadamente 20 ºC em alguns pontos no interior da bobina e do núcleo.

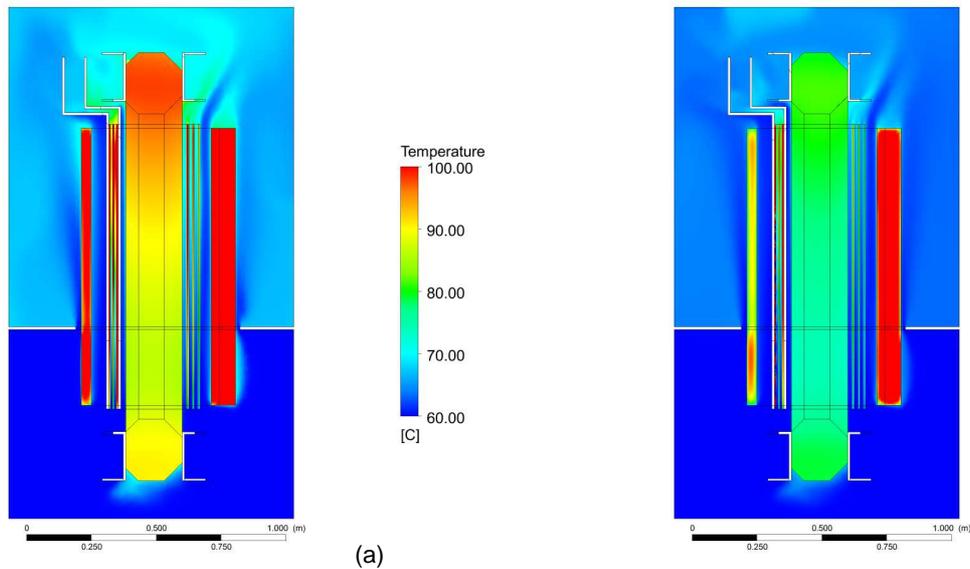

FIGURA 6 - Perfil de temperatura no plano de simetria. (a) modelo de 2 exaustores (b) modelo de 4 exaustores.

## 5. CONCLUSÕES

A utilização do modelo numérico permitiu avaliar e entender as características térmicas e fluidodinâmicas do escoamento no interior do transformador de energia.
Verificou-se que o aumento do número de exaustores promove um aumento significativo nos valores de velocidade no interior do transformador, em especial no interior dos canais entre os enrolamentos.
A utilização da placa central de vedação apresentou o efeito esperado, forçando o escoamento entre os canais dos enrolamentos e garantindo uma melhor refrigeração em ambos os modelos. As maiores temperaturas encontram-se nos enrolamentos centrais e na região superior.
A diferença no perfil térmico do ar na saída para os exaustores ficou em torno de 5°C entre os modelos com 2 exaustores e com 4.

Este tipo de análise fornece aos engenheiros de projetos uma maior confiabilidade na metodologia aplicada cotidianamente.

## 6. REFERÊNCIAS BIBLIOGRÁFICAS

## 7.  DADOS BIOGRÁFICOS

**Anderson Santos Nunes**

Nascido em Santa Maria, Rio Grande do Sul, em 08 de abril de 1986. Mestrando (2012) em Engenharia Elétrica pela Universidade Federal de Santa Catarina (UFSC), Pós Graduando (2011) em Projetos de Transformadores pela Universidade Regional de Blumenau (FURB) e Graduado (2010) pela Universidade Regional de Blumenau (FURB). Atua no setor de pesquisa e desenvolvimento em eletromagnetismo aplicado na empresa WEG T&D - Unidade de Blumenau.

**Marcelo Kruger**

Nascido em Ribeirão Preto, São Paulo, em 05 de novembro de 1980. Pós Graduando (2011) em Análise Mecânica através do Método de Elementos Finitos com ênfase em Aplicações Industriais pela empresa ESSS, Mestrado (2010) em Engenharia Mecânica pela Universidade Federal de Santa Catarina (UFSC), e Graduado (2004) pela Universidade Federal de Uberlândia (UFU). Atua como Coodenador de Contas na empresa ESSS. A empresa atua no mercado brasileiro e sul americano fornecendo produtos e serviços para empresas e instituições de P&D que demandam de soluções de simulação computacional de problemas de engenharia.